\documentclass[runningheads]{llncs}

\usepackage{tabularx}
\usepackage{csvsimple-l3}
\usepackage{rotating}
\usepackage{adjustbox}
\usepackage[english]{babel}
\usepackage{pgfplots}
\usepackage{hyperref}
\usepackage{color}
\usepackage{fancyvrb}

\usepackage{graphicx}%
\usepackage{multirow}%
\usepackage{booktabs}%
\usepackage[mode=image]{standalone}

\begin{document}

\title{PyRadiomics-cuda: 3D features extraction from medical images for HPC using GPU acceleration}

\author{Jakub Lisowski \and Piotr Tyrakowski \and Szymon Zyguła \and Krzysztof Kaczmarski}

\titlerunning{PyRadiomics-cuda: 3D features extraction using GPU acceleration}

\authorrunning{J. Lisowski et al.}

\institute{Warsaw University of Technology, Faculty of Mathematics and Information Science\\
            Corresponding author: \email{krzysztof.kaczmarski@pw.edu.pl}}

\maketitle


\begin{abstract}   
PyRadiomics-cuda is a GPU-accelerated extension of the PyRadiomics library, designed to address the computational challenges of extracting three-dimensional shape features from medical images. 
By offloading key geometric computations to GPU hardware it dramatically reduces processing times for large volumetric datasets. 
The system maintains full compatibility with the original PyRadiomics API, enabling seamless integration into existing AI workflows without code modifications. 
This transparent acceleration facilitates efficient, scalable radiomics analysis, supporting rapid feature extraction essential for high-throughput AI pipeline. 
Tests performed on a typical computational cluster, budget and home devices prove usefulness in all scenarios.
\end{abstract}


\section{Introduction and related works}

Radiomics extracts numeric features from medical images, turning images into analyzable data.
These features help build predictive models for cancer diagnosis, treatment response, and survival.
PyRadiomics is a widely used open-source Python library for radiomics \cite{van2017computational}.
It standardizes feature extraction from CT, MRI, and PET scans.
They include first-order stats, shape descriptors, and texture metrics (GLCM, GLRLM, GLSZM\footnote{Gray Level Co-occurrence Matrix, Gray Level Run Length Matrix, Gray Level Size Zone Matrix}).
Its modular design, 2D/3D support, and integration with tools like 3D Slicer \cite{slicer} and SimpleITK \cite{simpleitk} drove wide academic and clinical adoption.
Since release, PyRadiomics has been cited over 1700 times across many medical imaging studies.

However, shape-based features are computationally heavy, requiring 3D operations like mesh generation and pairwise distance calculations.
On large or high-resolution scans these steps are slow and limit scalability.
For example, mesh volume and surface area can be $O(n^3)$ and diameter calculations can be $O(m^2)$, with $n$ and $m$ the image linear resolution and ROI voxel count.
We found PyRadiomics can spend up to 99.9\% of time on diameter calculations for large ROIs (see Table \ref{tab:time-split}), a major bottleneck for large-scale AI workflows.

Several research groups have explored CUDA-enabled implementations to accelerate radiomics workflows, targeting the bottlenecks in shape and texture feature extraction. 
Rundo et al. presented a CUDA-based radiomics pipeline for unsupervised medical image analysis, demonstrating performance gains over CPU-based counterparts. 
Their approach emphasized parallel computation of texture features using CUDA kernels for high-throughput voxel-wise processing \cite{rundo_cuda-powered_2021}. 

There have been experimental forks of PyRadiomics incorporating GPU acceleration through PyCUDA or CuPy, particularly in research settings where feature extraction needed to scale across thousands of 3D scans. 
However, these implementations focused only on first-order feature classes (e.g., Grey Level Matrix), lacked full compatibility with the original PyRadiomics architecture (requiring significant restructuring of input pipelines), limited general usability, and are no longer maintained.

Another significant tool in this field is cuRadiomics \cite{jiao_curadiomics_2020}, an open-source library designed to replicate and extend PyRadiomics functionality on CUDA-enabled hardware. 
Unlike earlier GPU adaptations, cuRadiomics offers a comprehensive reimplementation of the radiomic feature extraction pipeline, leveraging NVIDIA’s CUDA architecture to accelerate a wide spectrum of feature classes, including first-order statistics, shape descriptors, and texture features such as GLCM and GLRLM. 
It incorporates memory-optimized GPU kernels, stream-based asynchronous execution, and a modular C++/CUDA architecture to support rapid feature extraction. 
In benchmarking studies, cuRadiomics has demonstrated speedups of 10× to 100× over traditional CPU-based methods. 
However, this performance gain comes with the trade-off of API divergence, as cuRadiomics does not maintain backward compatibility with PyRadiomics, often requiring tailored input formats and custom integration pipelines. 
It is also exclusively dedicated to the extraction of first-order statistical features (entropy, energy, etc.) and gray-level matrix features (contrast, entropy, and correlation of the matrix) and does not compute any geometric properties of the input images. 
Furthermore, cuRadiomics lacks meaningful integration with Python; it is merely an experimental implementation that could potentially be adapted for further use since running and configuring it is non-trivial. Its code repository has not been updated since 2021 and can hardly run with new CUDA and GPUs.

Unlike these prior efforts, PyRadiomics-cuda:
\begin{itemize}
    \item adds a CUDA backend that preserves PyRadiomics compatibility;
    \item auto-detects GPUs and falls back to CPU if no GPU is found;
    \item transparently accelerates 3D shape features. 
\end{itemize}

Obviously 3D features extraction is not the only bottleneck in radiomics workflows, with data reading and data transfer between CPU and GPU also contributing significantly to processing time.
In our research we analyzed the time spent on different steps of the workflow and justify PyRadiomics-cuda utilization in various configurations.

\section{PyRadiomics-cuda design}

From computational point of view, as it is visible in Table \ref{tab:time-split}  in case of large 3D volumes, PyRadiomics processing speed is limited by 3D features extraction time.
Therefore, PyRadiomics-cuda focuses exclusively on accelerating the \verb|Shape| feature class within PyRadiomics, which includes metrics such as: \emph{mesh volume, surface area, maximum 3D diameter and planar diameters (XY, YZ, XZ planes)}.

The process of finding the above volumetric features is based on two steps. 
Firstly it creates a mesh of triangles reproducing the 3D shape of the ROI in the image and on the fly calculates the volume and surface area of the shape.
In the second step it walks again the triangles in order to find the maximum 3D diameter and planar diameters.

\paragraph{Marching cubes fused parallel kernels.}

PyRadiomics uses a brute-force version of the Marching Cubes algorithm \cite{lorensen1987marching} for 3D mesh generation. 
It walks through all voxels in the 3D image, checking each voxel's eight corner values against a specified isovalue (typically the segmentation threshold). 
Number of voxels containing this isosurface is usually much smaller than the total number of voxels in the image, so this approach is efficient for smaller cases. 
Our parallel version of this process in general follows the same logic, but each voxel is processed in a separate thread. 
If a thread detects that the isosurface passes through its voxel, it computes the corresponding triangle and adds it to the global list of vertices and also updates the total volume and surface area.
Obviously this process requires synchronization between threads to ensure that the global data structures are updated correctly and here several possible approaches were tested (see Section \ref{sec:optimization}) in order to find the best one for different GPU architectures.

\paragraph{Parallel diameter calculation.}

Calculating the maximum 3D diameter of a shape is computationally intensive, as it finds the pair of vertices that are the farthest apart. 

The original PyRadiomics implementation suffers from inefficiency, especially for large meshes with many vertices.
To optimize this process for GPU execution, we again implemented a parallel algorithm that divides the workload among multiple threads. 
Each thread computes the distances between a subset of vertex pairs, storing the maximum distance found in an accumulator. 
After all threads complete their calculations, a reduction operation is performed to find the overall maximum distance from the local maxima. 

There are several ways to implement parallel accumulation process and we tested various strategies to find the best one for different GPU architectures (see Section \ref{sec:optimization}).

The resulting GPU kernels were optimized for performance and memory usage, leveraging shared memory, coalesced memory accesses to minimize latency and various parallel programming techniques \cite{cuda_handbook}.

\paragraph{PyRadiomics integration.}

The integration into the PyRadiomics ecosystem is designed to be seamless and transparent to the end-user. 
This is accomplished through a build-time injection mechanism that preserves full backward compatibility with the original API.

The process is managed by the convenient \verb|setuptools| configuration of the Python package, which has been extended to handle the CUDA compilation toolchain. 
During installation, the build system automatically detects the presence of the NVIDIA CUDA Compiler (\verb|nvcc|). 
If found, it compiles the C/CUDA source files of our extension. 
Before the compilation, within the C extension code, a single line calling the original CPU-based shape computation function is replaced with a call to the custom dispatcher function.
Later, at runtime, this dispatcher determines the execution path:
it first queries for a CUDA-capable device, if a suitable GPU is available and the driver initializes successfully, the computation is offloaded to the optimized CUDA kernels. 
If no GPU is found or an error occurs, the dispatcher gracefully falls back to the original CPU implementation, ensuring that the library remains functional on any system. 
This approach allows users to benefit from GPU acceleration without any changes to their existing code.
The usage is therefore the same as in the other PyRadiomics applications and is as simple as:
\smallskip
\begin{Verbatim}[fontsize=\scriptsize,numbers=left,numbersep=5pt,xleftmargin=0.3cm]
from radiomics import featureextractor
ext = featureextractor.RadiomicsFeatureExtractor()
res = ext.execute('scan.nii.gz', 'mask.nii.gz') #input 
print(res['MeshVolume'], res['SurfaceArea'])
\end{Verbatim}

\section{Optimization research and results}


\paragraph{Hardware used, testing data set and test framework.} 

There were three different machines used: (1) a professional modern GPU cluster built for AI, (2) an average desktop computer and (3) an old server machine equipped with a budget GPU designed to enable AI in refurbished hardware (specs listed in Table \ref{tab:configurations}). 
As an input data set we used public data from Kidney Tumor Segmentation Challenge 19 (KITS19) \cite{heller2020kits19,KidneyTumorSegmentation}. 
The complete set contains more than 200 cases, but we selected 20 samples covering exemplification of the possible input images sizes from 50kB to 9MB and producing from $2700$ to $236588$ 3D feature vertices. 

\begin{table}
    \caption{Configurations of the machines used for testing \label{tab:configurations}}
    \tiny%
    \begin{tabular*}{\textwidth}{@{\extracolsep\fill}llllll@{\extracolsep\fill}}
    \toprule
    No. & Machine Type & GPU Device & GPU Cores/VRAM & CPU Model & CPU Cores/Clock/RAM\\
    \midrule
    1 & Modern Cluster & NVIDIA H100 & 14 592 / 80 GB & AMD EPYC 9534 & 64 / 2.45 GHz / 1 TB \\
    2 & Desktop & NVIDIA RTX 4070 & 5 888 / 12 GB & AMD Ryzen 5 7600x & 6 / 5.3 GHz / 32 GB \\
    3 & Budget Cluster & NVIDIA T4 & 2 560 / 16 GB & Intel Xeon E5649 & 6 / 2.93 GHz / 18 GB \\
    \bottomrule
    \end{tabular*}
\end{table}






\paragraph{Micro-benchmarking the optimizations.}
\label{sec:optimization}

In order to find the PyRadiomics performance and the best possible combination of techniques that would yield the highest improvement for various GPU hardware and also rigorously validate the correctness of the results, we created a micro-benchmarking tool \cite{MiswutPyradiomicscudadatagen2026}.
It directly compares multiple algorithm implementations, including the original CPU baseline and numerous versions of the GPU kernels, each employing different optimization techniques.
All benchmarks use the same pipeline: reading the input images, applying the segmentation masks, transferring data to GPU (if needed), computing the features: volume, surface area and diameters.

Analysis of PyRadiomics in processing 3D shape features revealed that the main bottleneck appears in diameter calculation (see Table \ref{tab:time-split}), therefore we focused our optimization research on this part of the algorithm.

\begin{table}
\caption{\label{tab:time-split}The breakdown of the processing time in milliseconds into individual steps for PyRadiomics on CPU processor and PyRadiomics-cuda on GPU processor, both from desktop conf. 2 in Table \ref{tab:configurations}.
\textit{M.C.} -- Marching Cubes algorithm time, \textit{Diam.} -- 3D diameters calculation time, \textit{D. tran.} -- time needed to copy data from CPU to GPU memory, \textit{Comp.} -- speedup in pure computations, \textit{Overall} -- speedup including input file reading. 
Representative sample of KITS'19 data set \cite{heller2020kits19}.}
\centering
\tiny
\begin{tabular}{lcr|r|rrr|rrrr|rr}
\toprule
\bfseries case & \bfseries image & \bfseries vertices & \bfseries File &\multicolumn{3}{c|}{\bfseries PyRadiomics time}&\multicolumn{4}{c|}{\bfseries PyRadiomics-cuda time} & \multicolumn{2}{c}{\bfseries Speedup}  \\
\bfseries  no. & \bfseries size  & \bfseries in 3D & \bfseries reading & M.C. & Diam. & Total & D. tran. & M.C. & Diam. & Total & Comp. & Overall \\
 & & \bfseries space & [ms] & & & [ms] & & & & [ms] & & [x1]\\
\midrule%
00000-1 & 231x104x264 & 124406 & 2346 & 20.7  & 9516.5  & 9537.2  & 8.0  & 7.2  & 514.8  & 530.0  & 18.0  & 4.1 \\
00000-2 & 28x30x59 & 6132 & 2350 & 0.4  & 25.3  & 25.6  & 0.3  & 0.2  & 2.4  & 2.9  & 8.8  & 1.0 \\
00001-1 & 322x126x219 & 236588 & 2494 & 29.5  & 34210.3  & 34239.8  & 9.7  & 11.0  & 1855.8  & 1876.6  & 18.2  & 8.4 \\
00001-2 & 51x62x135 & 8928 & 2521 & 2.3  & 51.4  & 53.7  & 0.7  & 0.6  & 3.4  & 4.7  & 11.5  & 1.0 \\
00002-1 & 230x109x163 & 83098 & 1032 & 13.4  & 4256.2  & 4269.6  & 5.1  & 4.8  & 231.8  & 241.7  & 17.7  & 4.2 \\
00002-2 & 50x45x44 & 9206 & 1024 & 0.6  & 56.9  & 57.5  & 0.5  & 0.3  & 3.9  & 4.7  & 12.3  & 1.1 \\
00003-1 & 237x122x135 & 77560 & 1105 & 12.7  & 3731.0  & 3743.7  & 4.8  & 4.6  & 204.1  & 213.5  & 17.5  & 3.7 \\
00003-2 & 39x35x31 & 4568 & 1097 & 0.2  & 14.7  & 14.9  & 0.3  & 0.2  & 1.6  & 2.1  & 7.1  & 1.0 \\
00004-1 & 254x70x36 & 31838 & 254 & 2.5  & 677.2  & 679.7  & 0.8  & 1.1  & 37.8  & 39.7  & 17.1  & 3.2 \\
00004-2 & 35x37x10 & 2742 & 255 & 0.1  & 5.7  & 5.8  & 0.3  & 0.1  & 1.1  & 1.4  & 4.0  & 1.0 \\
00005-1 & 167x94x285 & 126446 & 3150 & 15.0  & 9780.9  & 9795.9  & 5.6  & 5.6  & 531.5  & 542.7  & 18.1  & 3.5 \\
00005-2 & 51x53x121 & 22024 & 3203 & 1.9  & 305.6  & 307.4  & 0.6  & 0.7  & 18.0  & 19.3  & 15.9  & 1.1 \\
00006-1 & 308x102x36 & 65436 & 710 & 4.4  & 2828.1  & 2832.5  & 1.1  & 2.0  & 153.7  & 156.8  & 18.1  & 4.1 \\
00006-2 & 41x43x13 & 3676 & 712 & 0.1  & 10.0  & 10.1  & 0.3  & 0.2  & 1.1  & 1.6  & 6.5  & 1.0 \\
00007-1 & 265x101x39 & 49912 & 255 & 4.1  & 1634.9  & 1638.9  & 1.0  & 1.7  & 90.1  & 92.8  & 17.7  & 5.4 \\
00007-2 & 39x43x12 & 3498 & 250 & 0.1  & 9.3  & 9.5  & 0.3  & 0.1  & 1.2  & 1.6  & 6.0  & 1.0 \\
00008-1 & 288x177x54 & 57362 & 967 & 9.3  & 2089.4  & 2098.6  & 3.3  & 3.1  & 113.7  & 120.1  & 17.5  & 2.8 \\
00008-2 & 127x154x41 & 47484 & 972 & 3.2  & 1436.9  & 1440.2  & 0.8  & 1.4  & 78.7  & 80.9  & 17.8  & 2.3 \\
00009-1 & 241x95x47 & 37576 & 337 & 3.8  & 916.2  & 920.1  & 1.1  & 1.5  & 50.5  & 53.0  & 17.4  & 3.2 \\
00009-2 & 39x33x11 & 2700 & 340 & 0.1  & 5.7  & 5.8  & 0.3  & 0.1  & 1.1  & 1.5  & 3.9  & 1.0 \\
\bottomrule
\end{tabular}
\end{table}

The initial implementation serving as a baseline (1) included basic techniques and equal threads load-balancing. 
The second optimization (2) added block-based atomic reductions. 
The third step improved theoretical memory access time (3) by using 2D structures loaded in the shared memory.
Since this optimization was not effective on older hardware with less shared memory per block, we introduced local thread accumulators (4) to reduce the number of atomic operations. 
Finally, we simplified the memory access patterns (5) to use just 1D arrays instead of 2D structures, which reduced the number of calculations needed for indexing. 
We found out that the conflicts between threads are rare, so the overhead of the simplest possible atomic operations is negligible.

\begin{figure}[ht]
    \centering
    \includegraphics[width=0.7\linewidth]{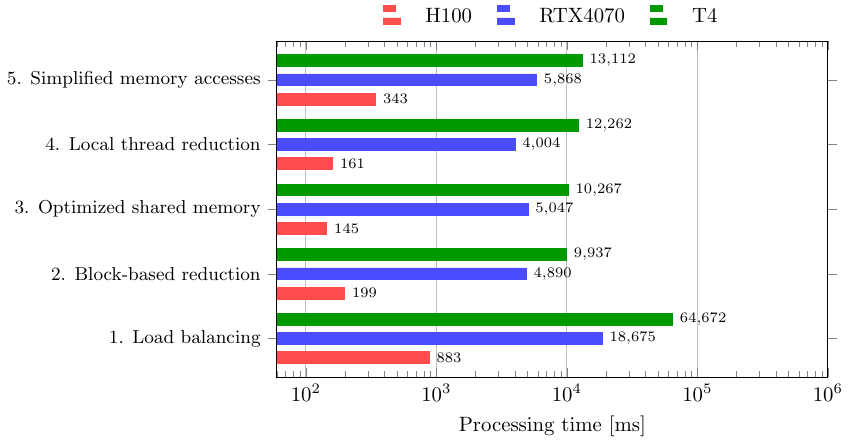}
    \caption{Comparison of optimization techniques applied to the GPU kernels for shape feature extraction across three GPU models. X-axis: sum of processing time of all input files in milliseconds on a logarithmic scale.  
    \label{fig:optimizations}}
\end{figure}

In Fig. \ref{fig:optimizations} we can see the comparison of the processing times for all the above-mentioned optimizations on tested GPU devices.
We observed that in case of an older T4 device atomic operations are not as effective and block-based reduction is the best strategy. 
Modern H100 offers fast atomic operations but needs more attention when accessing global memory. 
Finally, RTX4070 was the most effective with local thread accumulators.
Simplified memory accesses by using 1D arrays did not improve performance significantly on any device when compared to other ideas, so it was not included in the final implementation.

\paragraph{Discussion of the results.}

\begin{figure}[ht]%
\centering
\includegraphics[width=\linewidth]{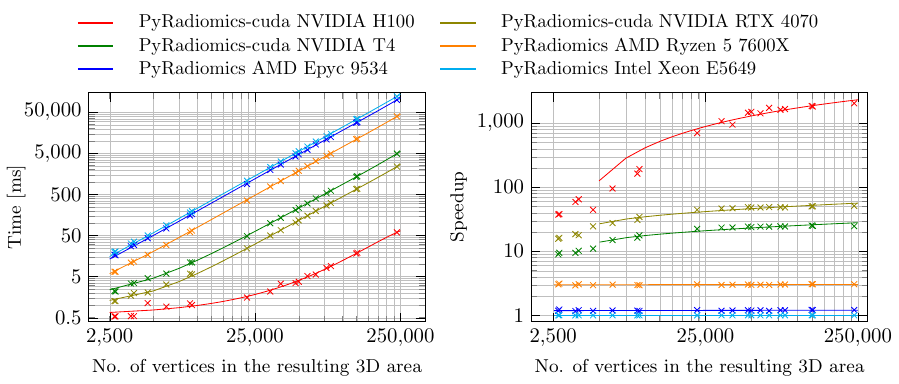}
\caption{%
\textbf{LEFT:}\ Time efficiency of the feature extraction using PyRadiomics and PyRadiomics-cuda on the various machines.\label{fig:results}
\textbf{RIGHT:}\ Available speedup of 3D features processing using PyRadiomics-cuda on various GPU devices compared to the original implementation working on CPU. Intel Xeon with PyRadiomics taken as a base reference time.\label{fig:speedup}}
\end{figure}

In Fig.~\ref{fig:results} we can see processing times of the KITS19 data set using PyRadiomics and PyRadiomics-cuda on various hardware configurations within a computational workflow including creation of a 3D shape from given images and ROI and then calculating volumetric features, diameters and area. 
The time is presented on log-log scale due to the large differences between the processing times for various datasets from 59 milliseconds on PyRadiomics-cuda on H100 GPU compared to 121 seconds for PyRadiomics on Intel Xeon on 236588 vertices and the same data.

A detailed analysis of the workflow in PyRadiomics indicated that most of the time after reading input file is spent in diameter calculation, from 95.7\% for small images to 99.9\% for large ones (see Table \ref{tab:time-split}). 
Reducing these times was the primary aim of the optimization.
Since PyRadiomics is not able to utilize multiple CPU cores, switching to better CPUs lead to only limited speedup, which could be attributed to the clock speed and architecture improvements.
This can be easily observed in Fig. \ref{fig:speedup}, where switching between CPU configurations 1, 2 and 3 did not reduce the processing time more than 3 times.
Using parallel processing on GPUs led to much higher speedup, which can be attributed to use of massively parallel threads. 
Even on a budget GPU, PyRadiomics-cuda achieved a speed-up from 8 to 24 times in 3D features extraction. 
For modern GPUs and more expensive hardware, the processing time was reduced by more than 50 and even 2000 times, compared to Intel Xeon processor.

In Table \ref{tab:time-split} we can also observe that PyRadiomics spends a long time loading data files. 
This cost comes not only from disk operations but also from various other operations like cleaning, normalization and conversion to proper memory layout.
Improving this step still remains an open challenge which could be done on GPU as well.
In all processed files, data transfer between CPU and GPU was compensated by faster processing and we did not observe any performance degradation. 
However, for smaller files the time spent on reading files was dominating the overall performance and no observable speedup was achieved.

For complete workflows (including data loading and preprocessing) performed on a desktop machine utilization of PyRadiomics-cuda resulted in 8 times speed-up for bigger data files with more than 200k vertices in 3D space. 
Direct Memory Access and fast data transfer will lead To huge savings in processing costs for machine learning tasks on clusters.

\section{Conclusions}

The initial motivation for this work was long time of feature extraction needed for training AI model in the xLUNGS project \cite{Luckner2025} dealing with approximately 40000 lungs CT (Computer Tomography) scans \cite{kolodko_polish_2025}, perhaps the world's largest chest X-ray database.
PyRadiomics-cuda significantly reduced 3D feature extraction time, leading to substantial savings in expensive infrastructure utilization, while delivering output with identical quality to the original PyRadiomics without requiring any changes to existing workflows.

We showed that PyRadiomics-cuda tackles the scalability and performance limitations of radiomics feature extraction, a core component of many biomedical imaging pipelines. 
By leveraging GPU acceleration while preserving compatibility with a widely adopted biomedical software framework, it bridges high-performance computing techniques and practical biomedical data analysis. 
The proposed solution enables efficient large-scale studies, supports reproducible research through open-source release, and provides a foundation for future extensions to distributed and multi-modal biomedical workflows.

PyRadiomics-cuda is implemented in Python and C/CUDA and as PyRadiomics is freely available under the BSD license at GitHub \cite{MiswutPyradiomicsCUDA2026}. 

\paragraph{Funding.} This research was carried out with the support of the High Performance Computing Center at Faculty of Math. and Inform. Science and POB Cybersecurity and data analysis of Warsaw Univ. of Techn. within the Excellence Initiative: Research University (IDUB) program.


\bibliographystyle{splncs04}

\bibliography{paper}

\end{document}